\documentclass[12pt]{article}
\pdfoutput=1

\usepackage{array}
\usepackage{amssymb}
\usepackage{graphics,graphpap}
\usepackage{graphicx}
\usepackage{color}
\usepackage{graphicx}
\usepackage{dcolumn}
\usepackage{epsfig}
\usepackage{epstopdf}
\usepackage{bm}
\usepackage{amsmath}
\usepackage{amsfonts}
\usepackage{textcomp}

\newcommand{\bea}{\begin{eqnarray}}
\newcommand{\eea}{\end{eqnarray}}

\newcommand{\bmp}{\noindent\begin{minipage}{16cm}}
\newcommand{\emp}{\end{minipage}\vskip 7mm} 

\def\drawbox#1#2{\hrule height#2pt
        \hbox{\vrule width#2pt height#1pt \kern#1pt
              \vrule width#2pt}
              \hrule height#2pt}

\def\Asym#1#2{\vcenter{\vbox{\drawbox{#1}{#2}
              \kern-#2pt 
              \drawbox{#1}{#2}}}}

\def\simge{\mathrel{%
   \rlap{\raise 0.511ex \hbox{$>$}}{\lower 0.511ex \hbox{$\sim$}}}}

\def\simle{\mathrel{
   \rlap{\raise 0.511ex \hbox{$<$}}{\lower 0.511ex \hbox{$\sim$}}}}

\def\s#1{\setbox0=\hbox{$#1$}%
\rlap{\ifdim\wd0>.7em\kern.22\wd0\else\kern.1\wd0\fi /}#1}

\def\be{\begin{equation}}
\def\ee{\end{equation}}

\def\c{\chi}

\def\g{\gamma}

\def\s{\sigma}
\def\t{\tau}

\def\z{\zeta}
\def\D{\Delta}

\def\L{\Lambda}
\def\O{\Omega}

\def\be{\begin{equation}}
\def\ee{\end{equation}}

\def\bea{\begin{eqnarray}}
\def\eea{\end{eqnarray}}

\begin{document}

\begin{titlepage}
\title{\vspace*{-2.0cm}
\bf\Large
Dark Radiation or Warm Dark Matter from long lived particle decays in the light of Planck}

\author{
Pasquale Di Bari\thanks{email: \tt P.Di-Bari@soton.ac.uk},~~
Stephen F.\ King\thanks{email: \tt S.F.King@soton.ac.uk},~~and
Alexander Merle\thanks{email: \tt A.Merle@soton.ac.uk}
\\ \\
{\normalsize \it Physics and Astronomy, University of Southampton,}\\
{\normalsize \it Southampton, SO17 1BJ, United Kingdom}\\
}
\date{\today}
\maketitle
\thispagestyle{empty}

\begin{abstract}
\noindent
Although {\it Planck} data supports the standard $\Lambda$CDM model, it still allows for the presence of Dark Radiation corresponding up to about half an extra standard neutrino species. We propose a scenario for obtaining a fractional ``effective neutrino species'' from a thermally produced particle which decays into a much lighter stable relic plus standard fermions. At lifetimes much longer than $\sim 1$ sec, both the relic particles and the non-thermal neutrino component contribute to Dark Radiation. By increasing the stable-to-unstable particle mass ratio, the relic particle no longer acts as Dark Radiation but instead becomes a candidate for Warm Dark Matter with mass ${\cal O}$(1~keV -- 100~GeV). In both cases it is possible to address the lithium problem. 
\end{abstract}

\end{titlepage}

\section{\label{sec:intro}Introduction}

The recently announced first cosmological {\it Planck} results~\cite{planck} herald a new era in cosmology in which the standard $\L$CDM model can be tested to high precision. For example, {\it Planck} results~\cite{planck} combined with WMAP polarisation and other CMB data measure the excess of radiation at recombination to be, in units of neutrino species,
\be\label{NeffplanckA}
N^{\rm Planck}_{\rm eff} = 3.36  \pm  0.34 \, ,
\ee
assuming a standard value of the primordial helium abundance. This result supports the standard $\Lambda$CDM model while not excluding the possibility of an extra radiation component, so-called Dark Radiation (DR), beyond the standard one. Indeed, when the {\em Planck} data are combined with the Hubble constant $H_0$ measurement from astrophysical data sets, in particular from the Hubble Space Telescope (HST), the best-fit value increases to
\be\label{NeffplanckB}
N_{\rm eff}^{\rm Planck} = 3.62 \pm 0.25  \,  ,
\ee
which amounts to a $2.3\,\s$ signal for DR. This shows that scenarios beyond the traditional (one fully thermalised) sterile neutrino hypothesis, in which a fractional ``effective neutrino species'' can emerge, are not excluded by {\it Planck} results and may even be mildly favoured by some data sets.\footnote{For a recent discussion (in the light of {\em Planck}) on models able to yield fractional ``effective neutrino species'', see~\cite{kaplan}. Notice that also in active-sterile neutrino oscillations, beyond a full sterile neutrino thermalisation, fractional ``effective neutrino species'' can be obtained~\cite{activesterile} and in this case the {\it Planck} data impose stringent constraints on the mixing parameters that seem to indicate a strong tension with the short-baseline hints~\cite{giunti}. A traditional solution, though difficult to justify, is to assume large initial lepton asymmetries suppressing the mixing~\cite{foot}.}

In this paper, motivated by the above considerations, we shall propose a scenario for obtaining a fractional ``effective neutrino species'' from a thermally produced particle which decays into a much lighter stable relic plus a non-thermal active neutrino component. On the other hand, by increasing the stable-to-unstable particle mass ratio to ${\cal O}$(0.1), the stable relic no longer acts as Dark Radiation but instead becomes a candidate for Warm Dark Matter (WDM). Thus our scenario is flexible enough to account for {\it either} DR {\it or} WDM (but not both at the same time). Interestingly, in both cases it is possible to address the lithium problem. However, before discussing details of our scenario, it is worth recalling some general constraints on new physics beyond the standard $\L$CDM model. Although well known, they are worth recalling at this point since they provide important constraints on our scenario.
 
Big Bang Nucleosynthesis (BBN) is the most traditional cosmological probe of new physics~\cite{shvartsman}. Non-standard BBN effects have been extensively studied within scenarios producing modifications of the neutrino content compared to the Standard Model (SM)~\cite{dolgov} such as a massive ($m_\nu = {\cal O}(10\,\rm MeV)$) decaying ordinary neutrino, now excluded by neutrino oscillation experiments~\cite{sato}, or active-sterile neutrino oscillations~\cite{activesterile}. On the other hand the inclusion of ordinary neutrino oscillations does not produce any modification to the standard scenario, in particular the SM value $N_{\rm eff}^{\rm SM}\simeq 3.046$ does not change~\cite{mangano}. Scenarios where massive metastable ($\tau \gtrsim 100\,$sec) particle decay products modify the abundances after nucleosynthesis have also been extensively investigated~\cite{lindley}.

However, the possibility to test non-standard BBN effects is greatly limited by systematic uncertainties in the determination of the primordial nuclear abundances~\cite{abundances}. The discovery of the acoustic peaks in the power spectrum of CMB anisotropies has opened new opportunities to probe non-standard effects with much lower systematic uncertainties. First of all, from CMB, it has been possible to measure with great accuracy and precision the baryon-to-photon number ratio $\eta_B$ that, if assumed to be constant between BBN and recombination time, allows to make firm predictions on the Standard BBN (SBBN) values of the primordial light element abundances to be compared with the measured values.

Moreover, from the observed acoustic peaks it is possible to constrain the presence of DR at the recombination time, the hot Dark Matter (DM) contribution, and even the primordial helium-4 abundance $Y_p$. Recent data from CMB observations have hinted to non-vanishing DR~\cite{data} and this has triggered quite an intense investigation on the possible sources, ranging from sterile neutrinos~\cite{Hamann:2010bk,Anchordoqui:2012qu} over modifications of the neutrino temperature~\cite{Boehm:2012gr} to exotic relativistic species~\cite{RelExotics}.

Non-standard effects from the decays of long-lived massive particles have been quite extensively investigated in general~\cite{scherrer}, in the case of inert product particles~\cite{scherrer2}, and in the case of electromagnetically interacting product particles, when they can alter the primordial abundances after nucleosynthesis~\cite{lindley}. In this case these have been advocated to reconcile a tension between the observed lithium abundance that is about three times lower than the value predicted by SBBN, the so-called lithium problem~\cite{lithiumproblem}.

Returning to our scenario, this needs to be considered carefully since it could potentially produce non-standard effects in a non-trivial way. For instance, the decay of a heavy thermally produced particle species into a new weakly coupled lighter stable relic plus non-thermal neutrinos, after the freeze-out of the neutron-to-proton abundance ratio and the neutrino decoupling, could alter the lithium abundance without affecting significantly the helium abundance. Concerning the details of our scenario, it is based on the particle physics model introduced in Ref.~\cite{King:2012wg}, where two Majorana fermions $\chi_{1,2}$ with masses $M_2 > M_1$ are added to the SM and both fermions can couple to the $Z$-boson, but only with vertices suppressed by factors $\epsilon_{1,2}$. However the cosmological implications of this model that we consider here are completely new. In particular the possibility for DR has not previously been considered, and the mechanism for WDM production here is quite different from the previous case where the $\chi_1$ was produced in thermal equilibrium. In the present case, only the $\chi_2$ is produced in thermal equilibrium and decays with a lifetime $\tau$ as $\chi_2 \to \chi_1 + f + \bar{f}$, where $f$ is any SM fermion, which (depending on the parameters) allows for {\it either} DR in $\chi_1$ and $\nu$ {\it or} WDM in $\chi_1$. DR emerges if the light stable relic $\chi_1$ is much lighter than $\chi_2$, while WDM is obtained by increasing the mass ratio to ${\cal O}(0.1)$.

The plan of the paper is the following. In Section~2 we discuss the basic features of the model. In Section~3 we calculate the DR contribution and show that this can explain the required value of $\D N_{\rm eff}\equiv N_{\rm eff} - 3.046$ from the combination of \emph{Planck} and HST data. In Section~4 we discuss the alternative scenario where the model provides an explanation for DM while the amount of DR would be negligible. In Section~5 we discuss how, in both cases, the lithium problem can be addressed, though for different ranges of values of the mass of the decaying particle. Finally, in Section~6, we draw the conclusions. In the Appendix we provide some technical details on the decay rate.

\section{\label{sec:model}The model}

Our setting extends the considerations of Ref.~\cite{King:2012wg}. We assume, in addition to the SM, two Majorana fermions $\chi_{1,2}$ with masses $M_{1,2}$. These fields are mainly SM-singlets, but have interactions with the $Z$-boson which are suppressed by factors $\epsilon_{1,2}$ and $\delta$. The interaction Lagrangians are given by
\begin{itemize}

\item $Z$--$\chi_i$--$\chi_i$ ($i=1,2$):
\begin{equation}
 \mathcal{L}_{ii} = g Z_{\mu}\epsilon_i^2 \overline{\chi_i} \gamma^\mu \gamma_5 \chi_i,
 \label{eq:Lii}
\end{equation}

\item $Z$--$\chi_1$--$\chi_2$:
\begin{equation}
 \mathcal{L}_{12} = g  \epsilon_1 \epsilon_2 \delta \overline{\chi_1} \gamma^\mu (A_\chi P_L + B_\chi P_R) \chi_2 Z_\mu + h.c.,    
 \label{eq:L12}
\end{equation}
\end{itemize}
($P_{R,L}=(1\pm \gamma_5)/2$) with $g\simeq 0.653$ being the generic $SU(2)$ gauge coupling and 
where $(A_\chi, B_\chi)$, with $A_\chi^2 + B_\chi^2 = 2$, 
are constants which parametrize the structure of the coupling. While at this stage these Lagrangians are only a postulate, similar settings are known to exist for example in the $E_6$SSM~\cite{E6SSM} or in certain Left-Right symmetric models~\cite{LR-models}.

The most important point is that such a setting admits a decay $\chi_2 \to \chi_1 + Z^*,\ Z^* \to f \overline{f}$, where $f$ is any SM-fermion. As long as the masses $M_{1,2}$ are suitably chosen, this in effect amounts to a transition $\chi_2 \to \chi_1 + f + \overline{f}$.
\footnote{As illustrated in Ref.~\cite{King:2012wg}, the other possible mode $\chi_2 \to \chi_1 W^+ W^-$ does not make much of a difference, since it is either kinematically forbidden or just a small perturbation. The mode $\chi_2 \to 3 \chi_1$ is even further suppressed. For simplicity, we neglect both of them.}
Then, we can make the simple observation that this reaction in particular allows for a decay involving light neutrinos
\begin{equation}
 \chi_2 \to \chi_1 + \nu + \overline{\nu}, \ \ \ {\rm if }\ M_2 > M_1.
 \label{eq:nu-dec}
\end{equation}
If we furthermore make the assumption that $\epsilon_2$ is large enough to keep $\chi_2$ in thermal equilibrium in the early Universe, while $\epsilon_1$ is sufficiently small that this does not happen with $\chi_1$, we can have the following phenomenologically interesting cases:
\footnote{Note that in the general setting presented here, there is no motivation for $\epsilon_1$ to be small other than to lead to DR. However, in Ref.~\cite{King:2012wg} some more concrete realizations of our setting are discussed, which do involve motivations for a small $\epsilon_1$. Alternatively, one can take the viewpoint that different parameter regions of the model simply lead to different interesting phenomenologies, which is another good motivation to explore them.}
\begin{enumerate}

\item Ordinary neutrinos and $\chi_1$'s as Dark Radiation:\\
In a certain parameter range we produce DR. Remarkably, a non-negligible fraction of this DR consists of \emph{ordinary neutrinos}, which might have further interesting implications. In other words, on top of the thermal neutrino component in the early Universe, the decays produce also a non-thermal contribution, similarly to a scenario discussed in~\cite{Cuoco:2005qr}. 

\item $\chi_1$ as Warm Dark Matter:\\
When a sufficiently large abundance of $\chi_2$-particles freezes out, they will all decay and each decay produces exactly one $\chi_1$. For the right combination of masses and couplings, $\chi_1$ could play the role of WDM if it is not too hot.

\end{enumerate}

We will now analyse both situations from a phenomenological point of view. It will turn out that indeed both points can be fulfilled in certain regions of the parameter space, however, they do not work out simultaneously. In other words, in the parameter regions where we obtain the correct WDM abundance we have practically no DR, and in the regions where we get reasonable amounts of DR the $\chi_1$'s also contribute to DR but not to WDM.

\section{\label{sec:DR_rec} The Dark Radiation scenario }

In this section we calculate $\D N_{\rm eff}$, defined as the value at recombination, within our model, to be compared with the value in Eq.~(\ref{NeffplanckB}), found by combining {\em Planck} data with the Hubble constant measurements. 

The total energy density $\rho_R$ in the radiation component receives a contribution from standard particles and from the $\chi_1$'s. In our case, if we restrict ourselves to the case where the helium abundance is standard, corresponding to setting $\tau$ to values much longer than the neutron-to-proton number ratio freeze-out time $t_{\rm fr}\sim 1\,$sec, the neutrino contribution can be unambiguously split into a standard thermal component $\rho_{\nu}^{\rm th}$ and into a non-thermal component $\rho_{\nu}^{\rm nth}$ resulting from the $\chi_2$ decays. The $\chi_2$'s are non-relativistic at the time of their decays and they have completely disappeared at the recombination time $t_{\rm rec}$. Therefore, the radiation contribution can be written as
\be
\rho_R(T) = g_R (T) \, {\pi^2 \over 30} \, T^4 \,  ,
\ee 
where the number of radiative degrees of freedom can be expressed as the sum of a SM component and of a non-standard component given by the non-thermal neutrinos and by the $\chi_1$ contribution so that $g_R(T) = g_{R}^{\rm SM}(T) + g_{\nu}^{\rm nth}(T) + g_{\chi_1}(T)$. The SM contribution is 
\be
g_{R}^{\rm SM}(T) = 2 + {7\over 8} \, \left[ g_{e^{\pm}}(T) + 
2\,\left({T_{\nu} \over T}\right)^4 \, N_{\rm eff}^{\rm SM}(T)\right] \,  .
\ee
At temperatures $T \ll m_{e}$ the contribution from $e^{\pm}$ vanishes, the neutrino-to-photon temperature ratio saturates to its asymptotical value $T_{\nu}/T = (4/11)^{1/3} \simeq 0.715$, and the effective number of neutrino species freezes to the value $N_{\rm eff}^{\rm SM} \simeq 3.046$, differing from $3$ since the thermal neutrino component is actually very slightly heated by $e^{\pm}$ annihilations. 

The non standard component can be analogously parametrised in terms of the extra number of effective neutrino species 
\be\label{DeltaNeff}
\Delta N_{\rm eff}(T \ll m_e) = [g_{\nu}^{\rm nth}(T)+g_{\chi_1}(T)]\,{4\over 7}\,
\left( {11 \over 4} \right)^{4\over 3}  =
{120 \over 7 \, \pi^2}\, \left( {11 \over 4}\right)^{4\over 3} \, 
{\rho_{\rm DR}(T) \over T^4} \,  ,
\ee
where we defined $\rho_{\rm DR} \equiv \rho_{\c_1} + \rho_{\nu}^{\rm nth}$. The energy densities of the non-thermal neutrino and $\chi_1$ components obey very simple fluid equations,
\begin{equation}
 \frac{d (\rho_\nu^{\rm nth} R^3)}{dt} = 
 \frac{b_{\nu}}{\tau} (\rho_{\chi_2} R^3) - (\rho_\nu^{\rm nth} R^3) H 
 \label{eq:Boltzmann_nuA}
\end{equation}
and 
\begin{equation}
 \frac{d (\rho_{\chi_1}\, R^3)}{dt} = \frac{b_{\chi_1}}{\tau} (\rho_{\chi_2} R^3) - 
 (\rho_{\chi_1}\,R^3) \, H,
 \label{eq:Boltzmann_nuB}  
\end{equation}
where $b_{\nu}=2\,{\rm BR}_{\nu}/3$ and $b_{\chi_1}=1/3$ are, respectively, the averaged fractions of energy into neutrinos and $\chi_1$ and ${\rm BR}_{\nu}$ is the branching ratio of $\chi_2$ decays into neutrinos. 

Notice that we are assuming $M_2 \gg M_1$ so that the $\chi_1$ can be treated as ultrarelativistic at the production. However, in order for the DR contribution not to be negligible, the $\chi_1$ have to be necessarily ultrarelativistic not only at the production but even until recombination since otherwise they would over-contribute to the DM energy density.
\footnote{This can be seen in a qualitative way imposing that the average momentum of $\chi_1$'s at $t_{\rm eq}$ (the matter-radiation equality time) $p_{\rm eq}\gg M_1$, leading to the condition $M_1/M_2 \ll 10^{-6}\,\sqrt{\t/{\rm sec}}$.} 
For this reason the $\chi_1$'s contribute to $\D N_{\rm eff}$ until recombination.

Assuming radiation dominance until equality, \footnote{Note that, in the region where we obtain a suitable value of $\Delta N_{\rm eff}$, $\chi_2$ can however never dominate the energy density of the Universe, as otherwise it would produce by far too much DR.} a solution of the differential equations is quite straightforwardly found,
\be\label{eq:solution}
\rho_{\rm DR} \, R^3 = b_{\rm DR}\,m_{\c_2}\,N_{\c_2}^{\rm f}\,\int_0^t \, dt' \, \frac{e^{-\frac{t'}{\t}}}{\t}\, \frac{a(t')}{a(t)} 
= b_{\rm DR}\,M_2\,N_{\c_2}^{\rm f}\, \sqrt{\frac{\t}{t}}\, \frac{\sqrt{\pi}}{2} \, \xi(t)  ,
\ee
having defined $b_{\rm DR} \equiv b_{\nu}+b_{\chi_1}$ and introduced
\be
\xi(t) \equiv  {\rm erf} \left( \sqrt{\frac{t}{\tau}} \right) - \frac{2}{\sqrt{\pi}} \, 
\sqrt{\frac{t}{\tau}} \, e^{-t/\tau} \,  ,
\ee
where the error function is defined as ${\rm erf}(x) = (2/\sqrt{\pi})\, \int_0^x e^{-z^2} dz$, such that simply $\xi(t) \stackrel{t \gg \t}{\longrightarrow} 1$. From Eq.~(\ref{DeltaNeff}) we can then calculate
\bea\nonumber
 \Delta N_{\rm eff}(t ) & \simeq & \frac{\zeta(3)\,45 \, \sqrt{2}}{7 \pi^{7/2}} 
 \,\left({8\,\pi^3 \over 90} \right)^{1\over 4}
 \left( \frac{11}{4} \right)^{4/3}\,d(t)\,g_{\chi_2} g_R^{1/4} \, b_{\rm DR}\,M_2 \,N_{\c_2}^{\rm f}\,
 \sqrt{\frac{\tau}{M_{\rm Pl}}} \, \xi(t)
 \\ \label{eq:Neff_rec}
& \simeq & 0.47 \, g_{\chi_2} \, d(t) \, b_{\rm DR}\,{M_2\,N_{\c_2}^{\rm f}\over {\rm MeV}}
\,\sqrt{\tau\over {\rm sec}}\, \xi(t)  \,  ,
\eea  
where $g_{\chi_2}=2$ is the spin degeneracy of $\chi_2$ and where we introduced the dilution factor $d(t) \equiv N_{\g}^{\rm f}/N_{\g}(t) $ and the $\chi_2$ relic abundance $N_{\c_2}^{\rm f}$ that got frozen at some freeze-out time $t_{\rm f}$. Notice that in the numerical expression we have approximated $g_R^{1/4} \simeq (g_R^{\rm SM})^{1/4}$, neglecting a small correction ($\simeq 2\%$) from the DR itself. We have also used a normalisation of the portion of co-moving volume $R^3$  such that the abundance of $\chi_2$ in ultrarelativistic thermal equilibrium is just $1$ (i.e.\ $N_{\chi_2}^{\rm eq}(T\gg M_2)=1$).

\begin{figure}[t]
\begin{tabular}{lr}
\includegraphics[width=8cm]{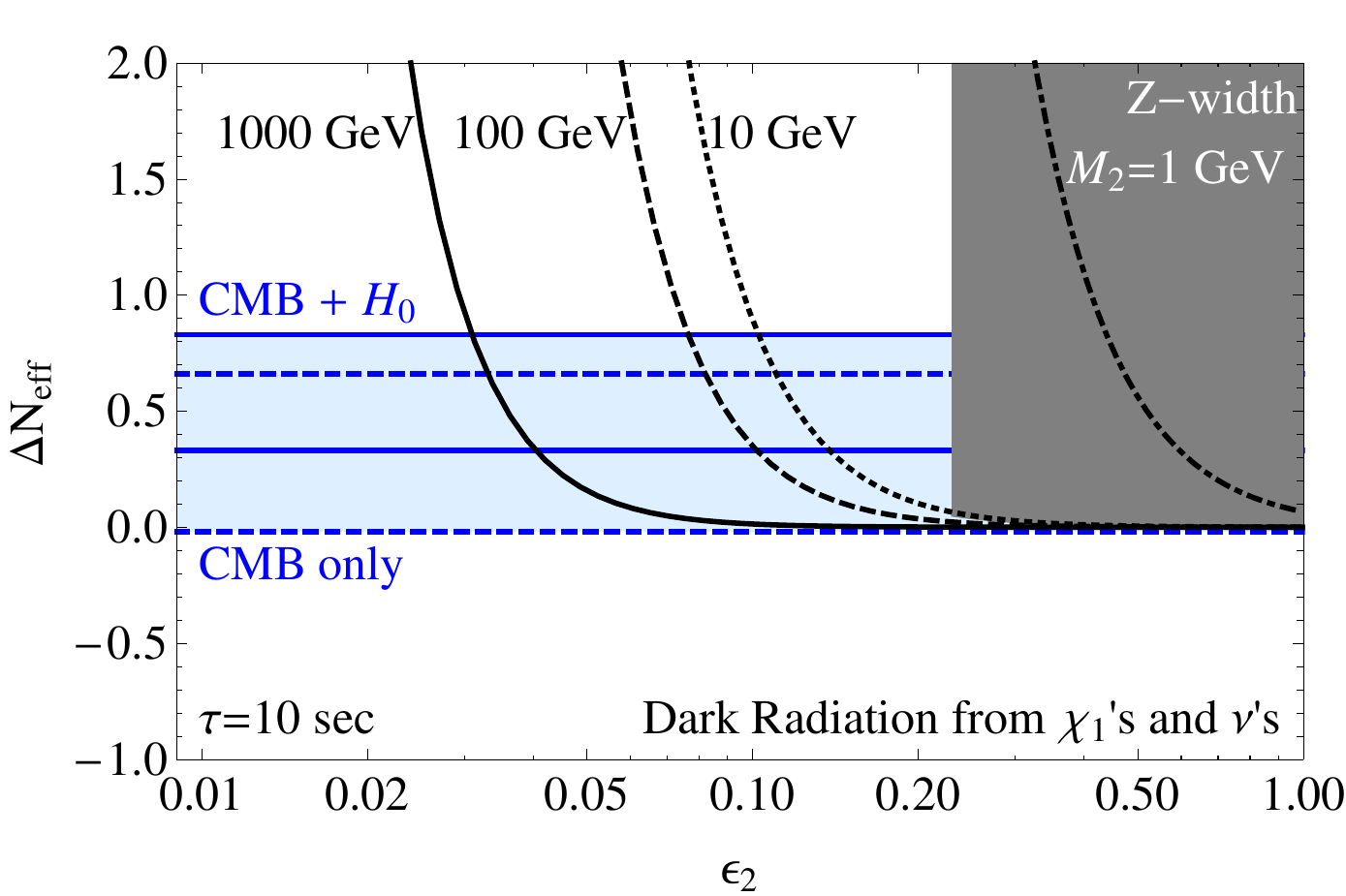} & \includegraphics[width=8cm]{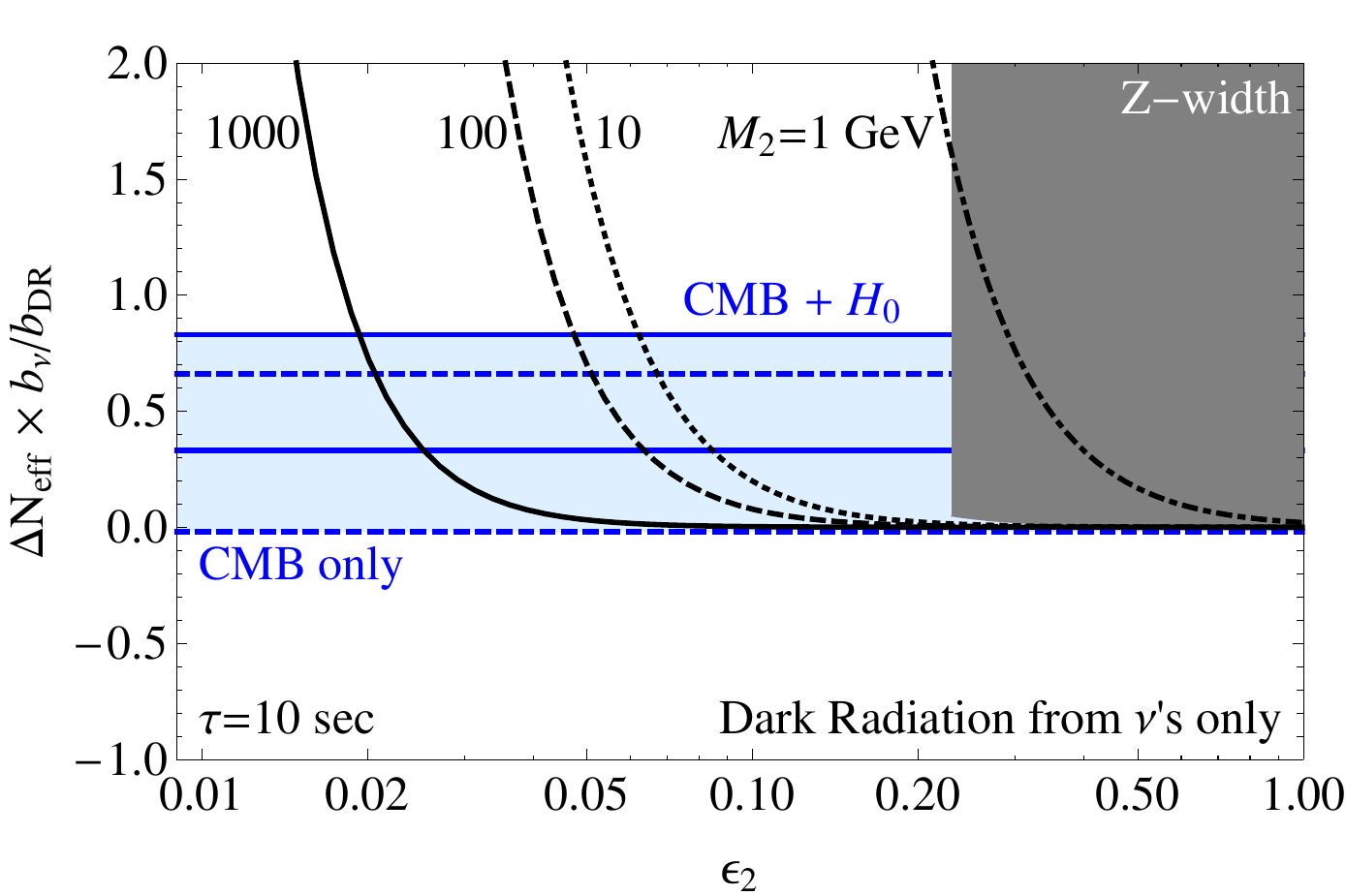}
\end{tabular}
\caption{\label{fig:Neff_CMB} Dark Radiation from $\chi_2 \to \chi_1 + f + \bar{f}$ at 
recombination time $t_{\rm rec}$ for a lifetime $\tau = 10$~sec.}
\end{figure}

The asymptotic value of Eq.~(\ref{eq:Neff_rec}) is then simply obtained by taking $\xi =1$ so that\footnote{This equation agrees with the result obtained in Ref.~\cite{Menestrina:2011mz} for decays just into inert particles, which would be formally recovered setting $b_{\rm DR}=1$ and taking into account that in our normalisation
\be
N_{\c_2}^{\rm f} = { Y_{\c_2}^{\rm f} \over d_0}\, 
\frac{g_{S0} }{ g_{\c_2}} \frac{ 8\,\pi^4}{ 135\,\z(3)}\, \,  
\ee
where $Y_{\chi_2}^{\rm f} \equiv n_{\chi_2}/s$, $s$ is the entropy density, $g_{S0}\simeq 3.91$ is the entropy number of degrees of freedom at the present time, and $d_0$ is the dilution factor at the present time. Notice that our model does also include this (more traditional) scenario that would be recovered for lifetimes $\t \ll t^{\nu}_{\rm dec} \lesssim t_{\rm fr}$, where $t_{\rm dec}^{\nu}$ is the time when standard neutrinos decouple. In this case one would then have simply $\rho_{\nu}^{\rm nth}=0$. For lifetimes $\t \sim t^{\nu}_{\rm dec}, t_{\rm fr}$ neutrinos would be in the decoupling stage and, therefore, this range would require a more complicated kinetic analysis both for the calculation of the contribution to $\D N_{\rm eff}$ from non-thermal neutrinos and, as we will point out in Section~5, also for the primordial helium abundance.}
\be\label{eq:Neffasymptotic} 
 \Delta N_{\rm eff}(t \gg \t)  \simeq  \Delta N_{\rm eff} \simeq
 0.47 \, g_{\chi_2} \, d_0 \, b_{\rm DR}\,{M_2\,N_{\c_2}^{\rm f}\over {\rm MeV}}
\,\sqrt{\tau\over {\rm sec}} \, . 
\ee  
In Fig.~\ref{fig:Neff_CMB} we show both the total contribution $\Delta N_{\rm eff} $ to DR (left panel) and just the neutrino contribution $(b_{\nu}/b_{\rm DR})\,\Delta N_{\rm eff} $ (right panel), for different values of the mass $M_2$ as functions of $\epsilon_2$. In the plots, we have $\epsilon_1 < 10^{-4}$, such that $\chi_1$ never enters thermal equilibrium and is only produced by $\chi_2$-decays. We have also chosen $\tau = 10$~sec for the plots, which can always be achieved by varying the remaining parameter $\delta$, cf.\ Appendix~A for more details.

The blue/light gray regions mark the $1\sigma$ regions from the \emph{Planck} plus other CMB data (dashed horizontal lines), as well as the corresponding regions when the $H_0$-measurements are also taken into account (solid horizontal lines), cf.\ Ref.~\cite{planck}. The dark gray region on the right is excluded by the invisible decay width of the $Z$-boson, which constrains $\epsilon_2$ to be smaller than about $0.23$ for $\epsilon_1 \ll \epsilon_2$, in case that $M_2 < M_Z/2$~\cite{King:2012wg}. As can be seen, for large enough masses $M_2$ we can always find a region of $\epsilon_2$ where we have a significant amount of DR while still satisfying the bounds. This contribution becomes large for small enough $\epsilon_2$, since this region corresponds to a very early freeze-out of $\chi_2$ and hence to an unsuppressed abundance.

Fig.~\ref{fig:Neff_CMB} might suggest a laboratory detection of the additional amount of ordinary neutrinos via neutrino capture on tritium or rhenium~\cite{NuCapt} or via (potentially resonantly assisted) modified electron capture on holmium~\cite{Lusignoli:2010eq,Li:2011mw}. Unfortunately, the rates turn out to be very small, due to the additional neutrino component in spite of the higher energy still being much less than the ordinary cosmic neutrino background. Even resonance enhancements~\cite{Lusignoli:2010eq} are not powerful enough to alter this conclusion.

\section{\label{sec:DM} Stable relic as Warm Dark Matter}

In Ref.~\cite{King:2012wg} it was shown that a thermal production of stable $\chi_1$'s would typically overclose the Universe for small masses $M_1$, but this can be cured by entropy production from $\chi_2$ decays to yield the keVin/GeVin scenario for DM. We take a different route here by assuming that $\epsilon_1$ is small enough ($< 10^{-4}$) to prevent $\chi_1$ from ever being in thermal equilibrium. We assume that the $\chi_1$'s are entirely produced from $\chi_2$ decays, so that we have a new way to produce a suitable DM candidate. Similar settings can be found in the literature, see e.g.\ Refs.~\cite{Kaplinghat:2005sy}. Since our scenario is more constrained than some of them, we have an important bound originating from the invisible $Z$-boson decay.

As had been shown in Ref.~\cite{King:2012wg}, one can thermally produce a non-relativistic abundance of $\chi_2$. \footnote{In principle, $\chi_2$ could also be relativistic at the production, but this would impose further complications and most probably not improve the result, since very hot particles would be produced.} 
The $\chi_2$ abundance is, as for a generic non-relativistic species, proportional to the mass $M_2$. However, $\chi_2$ is unstable and decays with a lifetime $\tau \gg t_{\rm fr}$. Since each $\chi_2$ decay produces exactly one $\chi_1$, this allows to simply translate the $\chi_2$ abundance into the abundance of the DM candidate $\chi_1$ (at the present time):
\begin{equation}
 \Omega_{\rm DM} h^2 = \Omega_{\chi_1} h^2 = \frac{M_1}{M_2} \Omega_{\chi_2} h^2,
 \label{eq:DM-conv}
\end{equation}
where $\Omega_{\chi_2} h^2$ is the final abundance of $\chi_2$ if it was the DM. By Eq.~\eqref{eq:DM-conv}, we can always correct an overabundance in $\chi_2$ such that the final DM abundance in $\chi_1$ hits the observed value, $\Omega_{\rm DM} h^2 = 0.1196 \pm 0.0031$~\cite{planck}.

Note that $\chi_1$ is produced in the early Universe at times $t \sim \tau$, but the expansion redshifts its momentum and, for this reason, even if it is ultra-relativistic at the production, it can still be slowed down by the cosmic expansion and be a viable DM candidate. We take this into account by calculating the free-streaming scale of $\chi_1$. This calculation can be found in textbooks (see, e.g., Ref.~\cite{Kolb:1990vq}), and it amounts to calculating the mean distance which the particle would travel if it was not trapped gravitationally. Technically, one has to evaluate:
\begin{equation}
 \lambda_{\rm FS} (t) = \int_{\tau}^{t} \, d\mathfrak{t}\, 
 \frac{v(\mathfrak{t})}{a(\mathfrak{t})} \,  ,
 \label{eq:FS-length}
\end{equation}
where $v(\mathfrak{t})$ is the velocity of $\chi_1$ and $a(\mathfrak{t})$ is the scale factor. Using elementary kinematics and the approximation of radiation-domination until equality, we obtain
\begin{equation}
 \lambda_{\rm FS} (t) \simeq 0.1\ {\rm Mpc}\ \sqrt{\frac{\tau}{10\ {\rm sec}}} \left( \frac{M_2/M_1}{10^2} \right)\, \ln \left( \sqrt{A} + \sqrt{1+ A} \right),
 \label{eq:FS-final}
\end{equation}
where 
\be
A \simeq 0.181\times 10^8\ \left( \frac{M_2/M_1}{10^2} \right)^{-2} \, \left( \frac{\tau}{10\ {\rm sec}} \right)^{-1} \,  .
\ee 
In order for the smallest structures in the Universe not to be erased, we need $\lambda_{\rm FS} \lesssim 0.1$~Mpc~\cite{Colin:2000dn}, and hence we must be in the region 
\be\label{condition}
{M_1 \over M_2}  \gtrsim 10^{-2}  \,  \sqrt{\frac{\tau}{10\ {\rm sec}}} \,  
\ee
(of course $M_1/M_2 < 1$). A free-streaming scale of $\sim 0.1$~Mpc corresponds to WDM, while values much below would correspond to cold DM. Therefore, for minimum values $\t \sim 10\,$sec one has already a marginal allowed region for cold DM ($M_1/M_2 \gtrsim 0.1$), that tends to disappear for longer lifetimes. For this reason in this setup DM is typically warm rather than cold.
\footnote{Notice, however, that in the case $\t \lesssim 10\,$sec, that we are not considering (cf.\ footnote~2), the cold DM region would become less marginal.}

\begin{figure}[t]
\centering
\includegraphics[width=10cm]{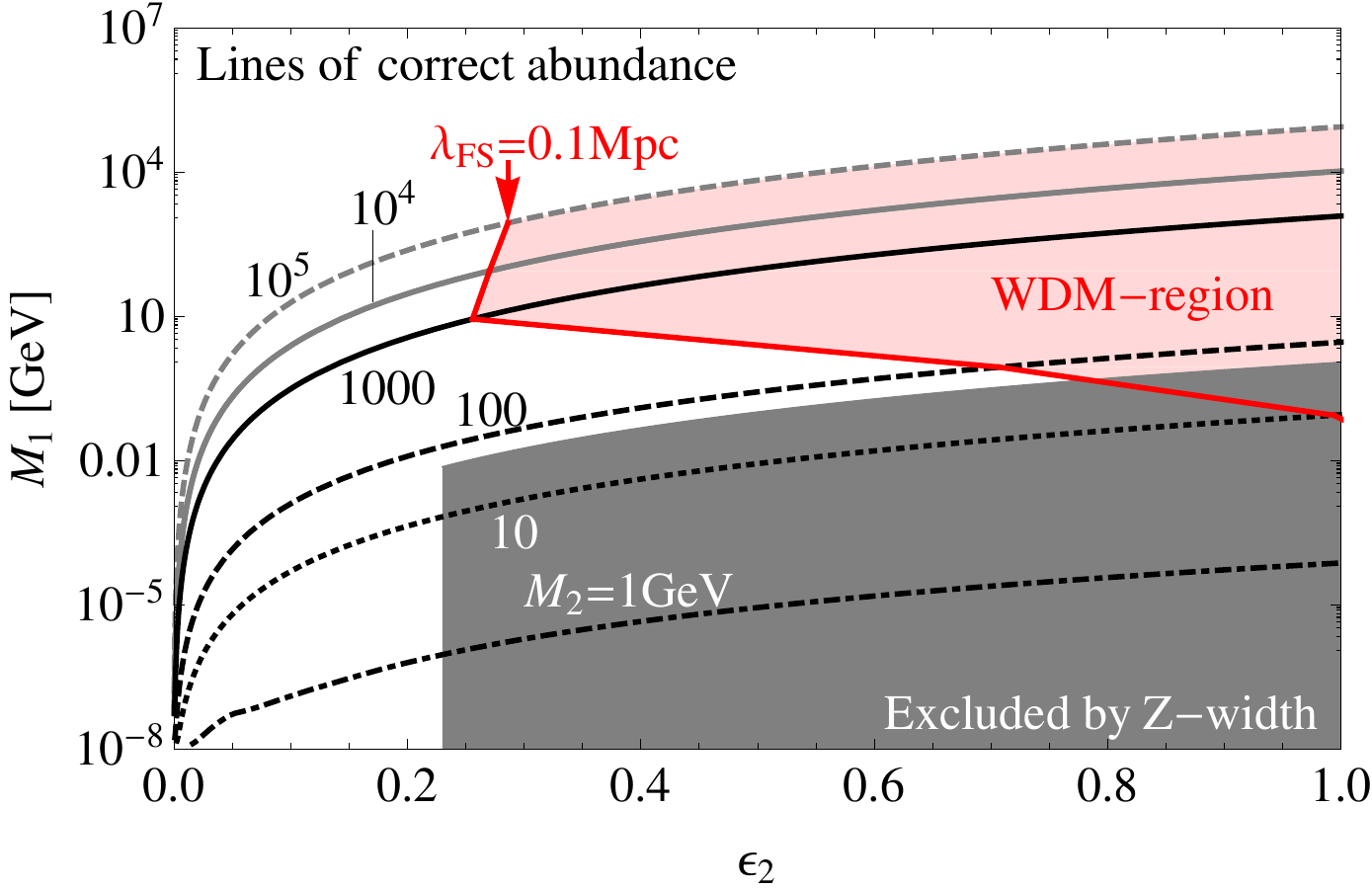}
\caption{\label{fig:M1req} The masses $M_1$ required to obtain the correct WDM abundance after the decay of all $\chi_2$'s (with $\tau = 10$~sec), displayed for different masses $M_2$ as functions of the suppression parameter $\epsilon_2$. The invisible $Z$-decay width requires $\epsilon_2 \lesssim 0.23$ for $M_2 < M_Z/2$. Note that we get \emph{warm} DM in the marked region, while CDM is only marginally possible due to $\epsilon_2 \leq 1$.}
\end{figure}

We have calculated the abundance of $\chi_2$ along the lines of Ref.~\cite{King:2012wg}, but with a more precise version of the decay width, cf.\ Appendix. The results can be seen in Fig.~\ref{fig:M1req}, where we have plotted the lines of correct abundance for different values of $M_2$ as functions of the suppression parameter $\epsilon_2$. Indeed, we are hit by the bound from $Z$-decay such that both $M_2$ and $M_1$ must be relatively large, which is very different from the scenario presented in Ref.~\cite{King:2012wg}. A particularly interesting point is that even a relatively heavy $\chi_1$ of $100$~GeV, or so, could be a \emph{warm} species produced at temperatures around $1$~MeV. The intuitive reason is that the energy that it obtains was in some sense ``stored'' at a relatively low temperature inside the non-relativistic $\chi_2$'s before they started decaying. Of course notice that the $\chi_2$'s do nevertheless decay much before matter-radiation equality and, therefore, they do not act as an additional DM component.

At this stage one could wonder whether the DM solution is compatible with having a sizeable DR contribution (calculated in the previous section), certainly an intriguing possibility. In this case the $\chi_1$'s would be already non-relativistic before matter-radiation equality and, therefore, only non-thermal neutrinos would contribute to DR. Unfortunately, as already anticipated at the end of Section~2, the model cannot lead to a sizeable amount of DR in the same region of parameter space where one has the correct $\lambda_{\rm FS}$, the region marked in pink/light gray in Fig.~\ref{fig:M1req}. This is easy to understand: a large value of $\epsilon_2$ keeps the $\chi_2$'s in equilibrium for a long time, such that their freeze-out abundance $N_{\chi_2}^{\rm f}$ is suppressed. While they can nevertheless produce a significant non-relativistic energy density in $\chi_1$'s, due to the large mass $M_1$, the mere amount of neutrinos produced by the reaction quoted in Eq.~\eqref{eq:nu-dec} is not enough to contribute significantly to DR given the observed value of $\O_{\rm DM}\,h^2$. This means that a sizeable value of $\Delta N_{\rm eff}\sim {\cal O}(0.1)$ is incompatible with the DM solution.

This can be easily understood also on quantitative grounds. Indeed if one plugs the DM condition Eq.~(\ref{eq:DM-conv}) into the Eq.~(\ref{eq:Neffasymptotic}), one finds the following relation linking $\D N_{\rm eff}$ to $M_1/M_2$,
\be
\D N_{\rm eff} \sim  10^{-3} \,
{\O_{\rm DM}\,h^2\over 0.1}\,{10^{-2}\over {M_{1}/ M_{2}} }\,\sqrt{\t\over 10 \, {\rm  sec}} \, .
\ee
One can then immediately see that, in order for the condition Eq.~(\ref{condition}) to be satisfied, one needs $\D N_{\rm eff} \lesssim 0.001$, saying that the DM scenario predicts, within conceivable experimental precision, a vanishing DR contribution.

Notice that there are two further restrictions on $\tau$: (i) it cannot be smaller than the time of the freeze-out of $\chi_2$ and also (ii) the condition $\delta \leq 1$ enforces $\tau$ not to be smaller than a certain minimum value depending (mainly) on $M_2$ (see the Appendix). Finally, on top of the further reaching cosmological aspects of our setting, we would like to add that one could even think of discovering our model at LHC, by directly producing $\chi_2$-pairs at high enough energies. We will postpone a more detailed investigation of all these interesting aspects to future work.

\section{\label{sec:lithium} Addressing the lithium problem}

The observation of acoustic peaks in the power spectrum of the CMB anisotropies gives a very precise measurement of the baryon-to-photon number ratio $\eta_B^{\rm CMB} =(6.030 \pm 0.075) \times 10^{-10}$~\cite{planck}. This can be used to derive quite precise predictions of the primordial nuclear abundances of light elements within SBBN. These can then be compared with the observed ones in astronomical environments. 

The primordial nuclear abundances that can be used as cosmological probes are the helium-4 abundance $Y_p$, the deuterium abundance D/H and the lithium abundance Li/H~\cite{PDG}. In the first case, with an (careful and significant) enlargement of the systematic uncertainties one finds, from clouds of ionized hydrogen (H{\sc ii} regions) in dwarf galaxies,
\be\label{Ypobs}
Y_p = 0.249 \pm 0.009  \hspace{5mm} (95\% \, {\rm C.L.})\,  ,
\ee
in agreement with the SBBN prediction~\cite{update} 
\be\label{YpSBBN}
Y_p^{\rm SBBN}(\eta_B^{\rm CMB}) \simeq 0.2466 + 0.01\,\log(\eta_{10}/5)\simeq 0.2474 \,  ,
\ee
where $\eta_{10} \equiv 10^{10}\,\eta_B$. 

The primordial deuterium abundance, measured from high-redshift, low metallicity quasar absorption systems, is found to be 
\be\label{deuterium}
({\rm D}/{\rm H}) = (2.82 \pm 0.21) \times 10^{-5} \hspace{5mm} (95\% \, {\rm C.L.})  \,   ,
\ee 
also in agreement with the inferred SBBN value 
\be\label{deuteriumSBBN}
{\rm (D/H)}^{\rm SBBN}(\eta_B^{\rm CMB}) \simeq 
3.6 \times 10^{-5}\,(\eta_{10}/5)^{-1.6} \simeq 2.7 \times 10^{-5}  \,  .
\ee 

On the other hand a stellar determination of the lithium abundance gives ${\rm (Li/H)}_p = (1.7 \pm 0.06 \pm 0.44) \times 10^{-10}$, about twice lower than the SBBN prediction ${\rm (Li/H)}_p^{\rm SBBN} \simeq (3-5)\times 10^{-10}$, the mentioned lithium problem that, barring still unknown conventional astrophysics mechanisms, provides a potential evidence of non-standard effects. 

It has been shown~\cite{lithiumproblem} that the decays (into hadrons and/or muons) of long-lived particles with a lifetime $\t = {\cal O}(10^4\;{\rm sec})$ and abundances ${\cal O}(1)-{\cal O}(100)$ times the baryon abundance, can reduce the lithium abundance by ${\cal O}(1)$ factors, thus solving the lithium problem, without spoiling at the same time the agreement of the deuterium and helium abundances.

These features can be easily fulfilled by our $\chi_2$ decays. If we impose the necessary conditions on the lifetime $\t$ and on the relic abundance $N_{\c_2}^{\rm f}$, we can show that we obtain physical solutions both in the DR and in the DM scenario.

Let us start with the DR scenario. First of all, the relic abundance $N_{\c_2}^{\rm f}$ is related to the $\chi_2$-to-baryon number ratio $\eta_{\chi_2}\equiv n_{\chi_2}/n_B$ at $\t \sim 10^4\,$sec by the relation $N_{\c_2}^{\rm f} \sim 10^{-8} \, \eta_{\chi_2}$. Therefore, plugging this relation into Eq.~(\ref{eq:Neffasymptotic}) with $\tau \sim 10^4$\,sec and imposing the {\em Planck} value $\Delta N_{\rm eff}\simeq 0.5$ (cf.\ Eq.~(\ref{NeffplanckB})), one finds easily $M_2 = {\cal O}(10\,{\rm GeV})-{\cal O}(1\,{\rm TeV})$. 

Let us now consider the WDM scenario. Repeating the same calculation with the condition $\D N_{\nu} \lesssim 0.001$ found in the WDM scenario, one obtains $M_2 = {\cal O}(100\,{\rm MeV})-{\cal O}(1\,{\rm GeV})$, having taken into account that necessarily $\chi_2$ should be heavier than muons for the solution of the lithium problem to be viable.

These results show that our model can nicely connect two independent cosmological puzzles, respectively, either DR or WDM and the lithium problem.

In the case of the DR scenario there is also another interesting consideration to be done. For interesting values of the lifetime $\t = {\cal O}(10^4\,{\rm sec})$, the model still predicts a SBBN value of $Y_p$ since $\Delta N_{\rm eff}(t_{\rm fr})$ would be still vanishing. Therefore, if future data supported values $\D N_{\rm eff}\simeq 0.5$, then it would be interesting to be able to test $\Delta N_{\rm eff}(t_{\rm fr}) \ll 0.5$, since in this way the model could be distinguished from a more traditional case where DR is generated prior to $t_{\rm fr}$ and then constant. The SBBN primordial helium abundance prediction would be modified by a quantity~\cite{update} $\D Y_p \simeq 0.0137\,\Delta N_{\rm eff}(t_{\rm fr})$. From Eq.~(\ref{Ypobs}) one then finds the bound $\Delta N_{\rm eff}(t_{\rm fr}) \lesssim 0.75$, that is still currently not stringent enough to provide an indication for such a scenario. It is interesting that CMB observations are also able to provide a measurement of $Y_p$, though currently the error is even larger than those from astronomical data sets Eq.~(\ref{Ypobs}). If future measurements will be able to constraint $\D Y_p \ll 0.006$ this could provide then another interesting piece of information.

Analogous considerations can be done for deuterium. In this case this is sensitive to $\D N_{\rm eff}(t_{\rm nuc})$, where $t_{\rm nuc}\simeq 365\,$sec is the time of nucleosynthesis. For life times $\t \sim {\cal O}(10^{4}\,{\rm sec})$ one would expect $\D N_{\rm eff}(t_{\rm nuc}) \ll \Delta N_{\rm eff}$, though in this case one could have some small contribution if $\t \sim {\cal O}(10^3\,{\rm sec})$. Then a non-vanishing $\D N_{\rm eff}(t_{\rm nuc}) $ would modify the SBBN into~\cite{update}
\be
{\rm (D/H)} \simeq {\rm (D/H)^{\rm SBBN}}(\eta)
\,[1+0.135\,\D N_{\rm eff}(t_{\rm nuc})]^{0.8} \,  .
\ee
Therefore, comparing the observed value Eq.~(\ref{deuterium}) with the SBBN prediction Eq.~(\ref{deuteriumSBBN}), one finds a constraint $\D N_{\rm eff}(t_{\rm nuc}) \lesssim 1.0$ that is also too large to draw a conclusion on a possible difference between $\D N_{\rm eff}(t_{\rm nuc})$ and $\D N_{\rm eff}$. However, a future improvement in the determination of ${\rm D/H}$ might make that possible if at the same time a non-vanishing $\D N_{\rm eff}\sim 0.5$ should be established. 

These considerations show how, in the case that future data should give an evidence for a non-vanishing $\D N_{\rm eff}\sim 0.5$, then a combined analysis with light element primordial abundances can potentially be used to highlight a dynamical evolution of DR as predicted by our model.
\footnote{This possibility was already pointed out and explored in earlier works, such as in
\cite{update} and in some of the papers in \cite{RelExotics}.}
From this point of view the current lithium problem would be interpreted as an effect of the $\c_2$-decays.

\section{\label{sec:conc}Conclusions}

We have seen that while {\it Planck} results support the standard $\Lambda$CDM model, they still allow for an amount of DR corresponding to about half an extra neutrino species, even favoured if data from astrophysical data sets (e.g. the HST) are taken into account. In particular, one could argue in support of cosmological non-standard effects that seem to require a dynamical mechanism for the production of DR at times after the freeze-out of the neutron-to-proton number ratio. 

Motivated by this situation, we have proposed a scenario for obtaining a fractional ``effective neutrino species'' from a thermally produced particle which decays, much after neutrino decoupling, into a much lighter stable relic plus a non-thermal active neutrino component.

Our scenario is based on two Majorana fermions, the heavier of which can decay into the lighter one plus a pair of ordinary fermions. If the lifetime is much longer than the neutrino decoupling time, the DR consists of two components, the lighter Majorana fermion and a non-thermal ordinary neutrino component and there would be an interesting, potentially testable, dynamical evolution of DR, with different values of $\Delta N_{\rm eff}(t)$ at the different observationally relevant times. For this reason we have limited our analysis to this range of lifetimes ($\tau \gg 1\,$sec). 

We found that a sizeable amount of DR, within the reach of future experimental investigations, is produced when the stable relic is orders of magnitude lighter than the decaying particles. On the other hand , by increasing the mass ratio $M_1/M_2$ to ${\cal O}(0.1)$, the stable relic no longer acts as Dark Radiation but instead becomes a candidate for WDM. In this way we obtained two mutually exclusive scenarios, one predicting a sizeable amount DR and one explaining DM.

Interestingly both scenarios, for $\tau \sim {\cal O}(10^4\,{\rm sec})$, are compatible with a solution of the lithium problem due to a partial disintegration of the synthesised value operated by the decay products if the mass of the decaying thermal relic particle is within the interesting range $M_2 = {\mathcal{O}(100\,{\rm MeV})}-{\cal O}(1\,{\rm TeV})$. Hence in both cases it is potentially possible to resolve the lithium problem. In the case of DR, this would also predict a SBBN and close-to-standard value respectively for the primordial helium and deuterium abundances (i.e. vanishing $\D N_{\rm eff}(t)$ at $t_{\rm fr}$ and $\D N_{\rm eff}(t)$ smaller than $\D N_{\rm eff}$ at $t_{\rm nuc}$).

In conclusion, we have seen that a relatively minimal extension of the Standard Model of particle physics is capable of predicting a few new signals in cosmology beyond the standard $\Lambda$CDM model. The proposed scenario seems to provide quite a flexible framework which could account for either DR or WDM. In the light of the first cosmological {\it Planck} data we have seen that there is a hint for DR and it is possible that a positive signal could be observed in future results. We eagerly look forward to the next results from {\it Planck} in the new era of precision cosmology in which scenarios such as the one presented here will be fully tested.

\section*{Acknowledgements}

PDB and SFK acknowledge financial support from the STFC Rolling Grant ST/G000557/1. SFK acknowledge financial support from the EU ITN grant UNILHC 237920. AM acknowledges financial support by a Marie Curie Intra-European Fellowship within the 7th European Community Framework Programme FP7-PEOPLE-2011-IEF, contract PIEF-GA-2011-297557. All three authors acknowledge partial support from the European Union FP7 ITN-INVISIBLES (Marie Curie Actions, PITN-GA-2011-289442).

\section*{Appendix: Decay width for $\chi_2 \to \chi_1 + \nu + \overline{\nu}$}
\label{app:ann}

\renewcommand{\theequation}{A-\arabic{equation}}
\setcounter{equation}{0}  

The decay width looks easiest for the case of the final state fermion $f$ being a neutrino, since then it is effectively massless. From Eq.~\eqref{eq:L12}, one can derive\footnote{For other SM fermions $f$ with mass $m_f > 0$ the expression looks more complicated.}
\begin{equation}
 \Gamma (\chi_2 \to \chi_1 \nu \bar{\nu}) = \frac{g^4 (\epsilon_1 \epsilon_2 \delta)^2 M_1}{2^7 \pi^3 M_2^2} \cdot A_\nu^2 \cdot [2 F_\chi \mathcal{\tilde I}_5 - H_\chi \mathcal{\tilde I}_3],
 \label{eq:decay_nu}
\end{equation}
where $\mathcal{\tilde I}_5 \equiv \int\limits_{t=0}^{M_2} dt\ g(M_2,M_1,0,t)\cdot t^2 \sqrt{1+ \frac{\Phi(M_2, M_1, t)}{4 M_2^2 M_1^2}}$, and $\mathcal{\tilde I}_3 \equiv \int\limits_{t=0}^{M_2} dt\ g(M_2,M_1,0,t)\cdot \frac{1}{2} t^2$. Furthermore, we have defined $A_\nu = \frac{g_V^\nu + g_A^\nu}{2 c_W}$, $F_\chi = A_\chi^2 + B_\chi^2$, and $H_\chi = 2 A_\chi B_\chi$, for an interaction Lagrangian $\mathcal{L} = g \tilde \xi \overline{\chi_1} \gamma^\mu (A_\chi P_L + B_\chi P_R) \chi_2 Z_\mu + h.c.$ Finally, the integrand functions $g(M_2, M_1, m_f, t)$ and $\Phi(M_2, M_1, t)$ are defined as
\begin{equation}
 g \equiv \frac{\sqrt{t^2 - 4 m_f^2} \sqrt{\Phi}}{(t^2 - M_Z^2)^2 + \Gamma_Z^2 M_Z^2},\ \Phi \equiv \left[ M_2^2 - (t - M_1)^2 \right] \left[ M_2^2 - (t + M_1)^2 \right].
 \label{eq:g_Phi}
\end{equation}
These are the expressions we have used to obtain our numerical results.

\end{document}